\documentclass[conference]{IEEEtran}
\IEEEoverridecommandlockouts
\usepackage{cite}
\usepackage{amsmath,amssymb,amsfonts}
\usepackage{algorithmic}
\usepackage[vlined,boxed,commentsnumbered,ruled,linesnumbered]{algorithm2e}
\usepackage{graphicx}
\usepackage{textcomp}
\usepackage{xcolor}
\usepackage{caption}
\usepackage{subcaption}
\usepackage{epstopdf}
\usepackage{booktabs}
\usepackage[colorlinks]{hyperref}

\newtheorem{definition}{Definition}
\newcommand{\lachesis}{\texttt{LACHESIS}}
\newcommand{\lachesisquad}{\texttt{LACHESIS }}
\def\BibTeX{{\rm B\kern-.05em{\sc i\kern-.025em b}\kern-.08em
    T\kern-.1667em\lower.7ex\hbox{E}\kern-.125emX}}
\begin{document}

\title{Learning to Optimize DAG Scheduling in Heterogeneous Environment}



\author{\IEEEauthorblockN{
Jinhong Luo\IEEEauthorrefmark{3}\IEEEauthorrefmark{2}\IEEEauthorrefmark{1},
Yunfan Zhou\IEEEauthorrefmark{2}\IEEEauthorrefmark{4}\IEEEauthorrefmark{1},
Xijun Li\IEEEauthorrefmark{2},
Mingxuan Yuan\IEEEauthorrefmark{2}, 
Jia Zeng\IEEEauthorrefmark{2}, 
and Jianguo Yao\IEEEauthorrefmark{3}, 
}
\IEEEauthorblockA{
\IEEEauthorrefmark{3}Shanghai Jiaotong University\\
\IEEEauthorrefmark{2}Huawei Noah's Ark Lab \\
\IEEEauthorrefmark{4}Chinese University of Hong Kong (Shenzhen)\\
\IEEEauthorrefmark{1}Equal contributions}
}

\maketitle
\footnote{Xijun Li is the corresponding author.}
\begin{abstract}
Scheduling job flows efficiently and rapidly on distributed computing clusters is one of huge challenges for daily operation of data centers. In a practical scenario, a single job consists of numerous stages with complex dependency relation represented as a Directed Acyclic Graph (DAG) structure. Nowadays a data center usually equips with a cluster of heterogeneous computing servers which are different in the hardware/software configuration. From both the cost saving and environmental friendliness, the data centers could benefit a lot from optimizing the job scheduling problems in the heterogeneous environment. Thus the problem has attracted more and more attention from both the industry and academy. In this paper, we propose a task-duplication based learning algorithm, namely \lachesis \footnote{The second of the Three Fates in ancient Greek mythology, who determines destiny.}, aiming to optimize the problem. In the proposed approach, it first perceives the topological dependencies between jobs using a reinforcement learning framework and a specially designed graph neural network (GNN) to select the most promising task to be executed. Then the task is assigned to a specific executor with the consideration of duplicating all its precedent tasks according to an expert-designed rules. We have conducted extensive experiments over standard workloads to evaluate the proposed solution. The experimental results suggest that \lachesisquad can achieve at most 26.7\% reduction of makespan and 35.2\% improvement of speedup ratio over seven strong baseline algorithms, including the state-of-the-art heuristics methods and a variety of deep reinforcement learning based algorithms.


\end{abstract}
\maketitle





\section{Introduction}



With the rapid development of parallel and distributed computing, scheduling job flows efficiently and rapidly on distributed computing clusters (i.e., data center) has become a vital problem. Because a little improvement in execution efficiency of the distributed computing clusters could both lead to huge increases in service providers' profits and large improvement of Quality of Service (QoS) for clients~\cite{little_change_big_money}. For this purpose, we are supposed to assign jobs to a suitable computation resource (also called executor) within the distributed system to complete jobs as soon as possible while meeting all requirements of service.

Specifically, in the typical jobs scheduling problem, a job to be scheduled consists of many stages (also called tasks) among which there is a determined dependent relationship. A Directed Acyclic Graph (DAG) is utilized to describe the above dependency. The job is said to be completed once its all tasks have been finished on executors following the given dependency. The goal is to finish on-the-fly jobs within a cluster of homogeneous executors with one or more metrics of interest (e.g., makespan, average complete time, etc.) via assigning each task to an appropriate executor subject to all dependencies between tasks. The job scheduling problem within homogeneous executors has been proved NP-hard~\cite{np_hard}. However, the heterogeneous computation has become the mainstream infrastructure of cutting-edge distributed computing clusters, which requires us to consider the differences among the executors when scheduling jobs to these heterogeneous executors. The differences between these heterogeneous executors might be the computation, communication capacity, etc., which greatly increases the complexity of the scheduling problem.

Many previous works have been proposed to solve the DAG scheduling problem under different settings, which fall into two categories, \textit{heuristics}, and \textit{learning-based algorithms}. Generally, heuristics mainly include list scheduling~\cite{cpop_heft}, cluster-based scheduling~\cite{dsc}, and task duplication-based scheduling~\cite{schedule_two_phase}. List scheduling~\cite{cpop_heft} divides the scheduling process into two-phase work, task selection, and executor allocation. Cluster-based scheduling~\cite{dsc} clusters the tasks by heuristic rules to resolve the dependencies and assigns different task clusters to specific executors by the assignment strategy. Task duplication-based scheduling~\cite{tds} duplicates tasks properly to reduce communication costs to minimize the completion time of all jobs. Furthermore, the task duplication technique can be combined with list scheduling or cluster-based scheduling to improve the scheduling result. 
However, the design of heuristics heavily relies on the experience of human experts, almost not taking advantage of amounts of redundant pattern and characteristics of the scheduling problem.

Thus, learning-based algorithms are proposed to address the above issues. Mao \textit{et. al.} proposed Decima algorithm~\cite{decima} that adopts a Graph neural Network (GNN) to learn the task dependencies and executor real-time information first and then decides the stage assignment. Besides, Sun \textit{et. al.} proposed a GNN-based architecture named DeepWeave\cite{sun2021deepweave}, which reduces the Job Completion Time (JCT) by accelerating the co-flow transmission while processing every single job stage. Nevertheless, all above techniques neglect the development of heterogeneous computation, which means they do not consider the DAG job scheduling within a cluster of heterogeneous executors. 

Thus, to optimize the DAG Job scheduling problem, \lachesis is proposed in this paper. The major challenges is summarized as follows. First, we have to tackle the complex dependencies among job stages, which requires the algorithm to comprehensively perceive the topological structure of jobs. Second, the state information in a scheduling problem is large-scale and dynamic, since hundreds of jobs are continuously submitted by clients to the distributed system. Third, the heterogeneity of executors makes the calculation of an optimal solution very difficult. To address the above challenges, a task-duplication-based learning algorithm named \lachesis~ is proposed, which incorporates GNN with an elaborated heuristic in a two-phase scheduling framework. In the first phase, \lachesis~ perceives the dependencies between all existing tasks using a specially designed GNN, followed by a policy network to select one task to be executed. Then a heuristic method is designed to assign the task to the most 'appropriate' executor and meanwhile decide whether to duplicate all its precedent tasks or not.

The technical contributions of this paper can be summarized as follows:
\begin{itemize}
\item We define a DAG job scheduling problem in a heterogeneous environment, where the executors might differ from each other with respect to capabilities of computation and communication. Meanwhile, the task to be executed in the environment is allowed to be duplicated over multiple executors.

\item We design a task-duplication based learning algorithm, which incorporates the GNN with heuristics in a two-phase framework. The two phases are in charge of task selection and executor allocation respectively. Specifically, in the first phase, we adapt the GNN in order to deal with DAGs of arbitrary shapes and sizes and complex dependencies. Besides, the heuristics used in the second phase consider the duplication of tasks over multiple executors.

\item We evaluate the proposed algorithm with extensive experiments. The experimental results show that our approach achieves at most 26.7\% reduction of makespan and 35.2\% improvement of speedup ratio over seven strong baselines.

\end{itemize}

The remainder of this paper is organized as follows. The previous works are discussed in Section \ref{related_work}. A formal definition of the problem is given in Section \ref{definition}. Our approach to solving the problem is presented in Section \ref{design}. The experimental results are reported in Section \ref{evaluation}, followed by conclusion in Section \ref{conclusion}.


\section{Related work}\label{related_work}

Many works has been contributed to solve the DAG scheduling problem with both heuristic and machine learning methods. Domain Resource Fairness (DRF)~\cite{fair_scheduling} algorithm schedules different types of resources to satisfy several highly desirable properties, especially fairness. The Shortest Job First (SJF) algorithm schedules the shortest job preferentially. Tetris~\cite{tetris} schedules many types of executors with an analog of shortest-running-time-first to trade-off cluster efficiency for speeding up individual jobs.

List scheduling is widely used for DAG scheduling. Heterogeneous Earliest Finish Time (HEFT) algorithm~\cite{cpop_heft} is a highly efficient list scheduling algorithm. Specifically, HEFT prioritizes the task with a high-level feature and allocates executors for tasks with a heuristic search algorithm. Besides, Critical Path on a Processor (CPOP) algorithm~\cite{cpop_heft} is another list scheduling algorithm which allocates the fastest executors for the tasks on critical path and matches other executors with other tasks. Moreover, the Dynamic Level Scheduling (DLS) algorithm~\cite{dp_scheduling} chooses a pair of tasks and executors which maximize a dynamic attribute combined with tasks and executors, which is also a list scheduling algorithm.

The scheduling policy is able to duplicate some tasks to save time for data communication in task duplication scenario. Task Duplication based Scheduling (TDS) algorithm~\cite{tds} uses clustering scheduling method to generate a schedule only considering duplicating the critical parent task. Duplication First and Reduction Next (DFRN) algorithm~\cite{dfrn} based on list scheduling method duplicates the task firstly and reduces the redundant task secondly. Task Duplication based Clustering Algorithm (TDCA)~\cite{tdca_algorithm} generates a schedule through four steps: cluster initialization, task duplication, process merging, and task insertion.

Different from the aforementioned heuristic algorithms, learning-based algorithms including Reinforcement Learning (RL)~\cite{RL1998}, Deep Learning (DL)~\cite{deep_learning} etc. are proposed to solve DAG scheduling problem. Decima~\cite{decima} using RL to train a deep neural network to schedule the jobs on spark cluster. However, Decima only handles the scheduling in homogeneous environment and detachable jobs without considering data transmission time existing in jobs. Furthermore, task duplication is not supported by Decima. DeepWave~\cite{deepwave} proposed by Sun \textit{et. al.} also uses RL to train a neural network to schedule coflow in job DAG in order to reduce Job Complete Time (JCT). Ni \textit{et. al.} proposed Dream~\cite{dream} which places a data processing node on a suitable resource to maximize the throughput of the whole process using GNN and RL.


\section{Problem Description}\label{definition}




DAG scheduling problem in heterogeneous environment widely exists in distributed computing systems. The jobs to be executed are continuously submitted by clients to the distributed system which is composed of heterogeneous executors. In particular, each job consists of tasks, in which there are determined dependencies. One task could be executed on one executor if 1) all its precedent tasks have been finished and 2) corresponding data needed is available. A job is said to be completed if all tasks of it have been finished by corresponding executors. One is supposed to appropriately assign tasks to executors in order to complete on-the-fly jobs as soon as possible while meeting all requirements of scheduling. Specifically, the goal of this work is to minimize the makespan. A relatively formal problem description is given below. For ease of reading, we summarize all variables used in this paper in Table~\ref{tab:notation}.

\begin{table}[t]
 \caption[Notation of all variable used in this paper]{Notation of all variable used in this paper}
 \label{tab:notation}
 \centering
 \begin{tabular}
 {l | c@{} }
  \toprule
  symbol   & description\; \\
  \midrule
  $J$    & job (DAG)\; \\
  $n$    & task (DAG node)\; \\
  $w_i$   & the computation size of node $n_i$\;  \\
  $j(n_i)$   & node $n_i$'s job\; \\
  $\varphi(n_i)$ & node $n_i$'s parent nodes\; \\
  $\xi(n_i)$ & node $n_i$'s children nodes\; \\
  $\gamma(n_i)$ & edges connecting node $n_i$ and its children nodes\; \\
  $\mathcal{A}_{t}$ & the set of executable nodes at time $t$ \\
  $e_{ij}$  & the size of data transferred from node $n_i$ to node $n_j$\;  \\
  $r_k$   & the executor in system\; \\
  $v_k$   & the processing speed of executor $r_k$\;  \\
  $R_{n_p}$ & the set of executors allocated for node $n_p$\;  \\
  $q_n^{j}$ & node score\;  \\
  $c_{ij}$  & the transmission speed between executor $r_i$ and executor $r_j$\; \\
  $f,g,q$  & non-linear functions\; \\
  $x_n^{j}$ & node feature vector\; \\
  $a_n^{j}$ & edge feature vector\;\\
  $e_n^{j}$ & per-node embedding\; \\
  $EST$   & the earliest start time\;  \\
  $AST$   & actual start time for job (node)\;  \\
  $EFT$   & the earliest finish time\;  \\
  $AFT$   & actual finish time for job (node)\;  \\
  \bottomrule
 \end{tabular}
\end{table}

\begin{definition}[Actual Finish Time]
$AFT(n_i, r_k)$ represents actual finish time for node $n_i$ executed on executor $r_k$, which can be calculated as follows:
 \begin{equation}
  \label{eq:aft}
  AFT(n_i, r_k) = AST(n_i, r_k) + \frac{w_i}{v_k}
 \end{equation}
where $AST(n_i, r_k)$ denotes actual start time for node $n_i$ is executed on executor $r_k$.
\end{definition}

\begin{definition}[Earliest Start Time]
$EST(n_i, r_j)$ represents the earliest start time for node $n_i$ executed on executor $r_j$, which can be calculated by:
 \begin{equation}
  \label{eq:est}
  EST(n_i, r_j) = \max_{n_p\in \varphi(n_i)} ( \min_{r_p \in R_{n_p}} AFT(n_p,r_p)+\frac{e_{p,i}}{c_{p,j}})
 \end{equation}
 \begin{equation}
  \label{eq:est_eft}
  EFT(n_i, r_j) = EST(n_i, r_j) + \frac{w_i}{v_j}
 \end{equation}
\end{definition}

Obviously, the earliest start time and the earliest finish time depends on actual finish time of parent nodes and dependent data transmission time.

\begin{definition}[Makespan]
Makespan is completion time of the workload which is significant to execution efficiency of the whole process. Our target in the DAG scheduling problem is minimizing the makespan, which can be defined as follows:
\begin{equation}
  \label{eq:objective}
  Objective: \min(makespan(Job))
\end{equation}
where $Job = \{J_1,J_2,\cdot\cdot\cdot,J_k\}$ represents the set of jobs in the system.
\end{definition}

\begin{figure}[h!]
 \centering
 \includegraphics[width=0.88\linewidth]{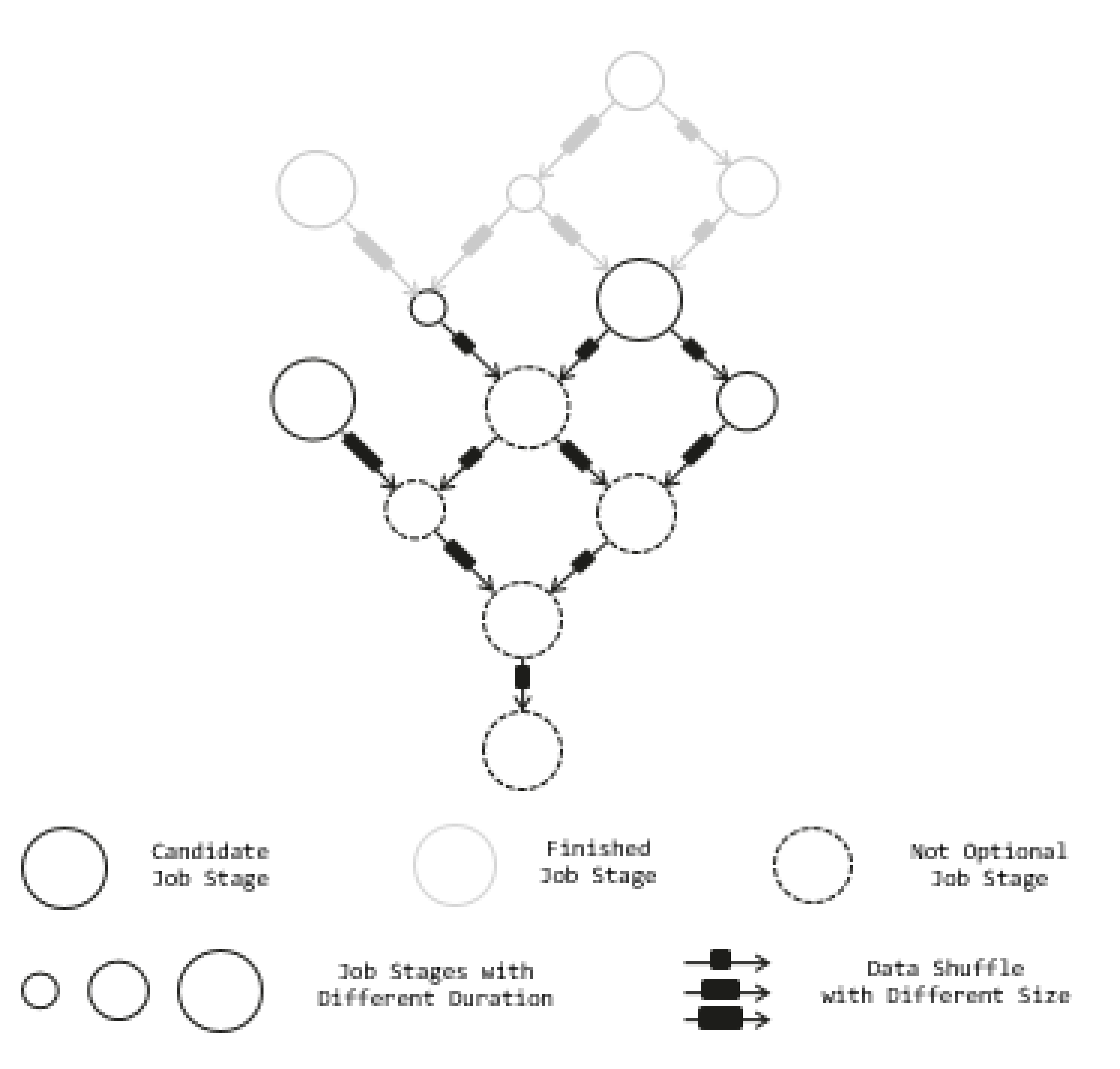}
 \caption{A simple instance of a DAG-structured job. A node in the DAG represents a job stage, which is composed of several parallel tasks. Meanwhile, an edge is defined by the nodes it connects and the size of data shuffled by it. In a DAG job, only the nodes whose parent nodes are finished can be the candidate of the next node chosen to be processed.}
 \label{fig:whole_abstract_algorithm}
\end{figure}

The DAG scheduling problem can be illustrated as follows. Given the dependency relations of tasks, the computation size of the task, the size of data transferred between tasks, speed of computation for executors, and speed of data transmission for executors, the DAG scheduling is to schedule the tasks on executors to minimize the makespan. In order to simplify the problem properly, we assume that every task can be executed by executors successfully without considering execution failure. Furthermore, we assume that $\frac{w_i}{v_j}$ represents all time for task $n_i$ executed on executor $r_i$ avoiding considering the changes of the processing speeds of executors. In addition to assumptions, the constraints of the problem are also significant to the scheduling process. And some constraints must be satisfied in scheduling process. The constraints can be demonstrated as follows:


1) \textit{Constraints for tasks:} First, all tasks are atomic, and they can't be preempted by another task if they are running on executors. Moreover, a task should be assigned to at least one executor to be executed. If it is assigned to more than one executor, the result of the same task executed on different executors is also the same. The last, a task can be executed by an executor until all of its dependencies are satisfied and all data required is transferred to the assigned executor. Finally, tasks can be noticed and processed if and only if the corresponding job arrives at the system.

2) \textit{Constraints for executors:} First of all, $|R|\in \mathcal{N} $ indicates that the number of executors in the system is limited. Then, every executor can only execute a task at the same time. In other word, the execution of different tasks assigned on the same executor is exclusive. Finally, $v_i \neq v_j, i\neq j$ represents that the processing speeds of executors may be different.

3)\textit{Constraints for communication:} In the first place, result data can only be transferred from one executor to another executor after the task generating result data is completed. In the second place, data transfer time is communication data size divided by transfer speed between two executors if the parent task and the children task aren't assigned to the same executor, and otherwise 0.

The main challenges of DAG scheduling in a heterogeneous environment could be summarized as follows.

\noindent\textbf{Complex dependencies.}
Data processing platforms and query compilers such as Hive, Hadoop, and Spark-SQL create jobs which usually have a DAG structure. Generally, the DAG is complicated and arbitrary due to tens or hundreds of tasks and countless data transmissions. An optimal scheduler should ensure that the tasks are executed in parallel as much as possible and no task is blocked by the dependencies of the task under the condition of available executors. In order to realize the above targets, the scheduling algorithm should understand the complex dependencies of tasks and plan ahead to assign the tasks to executors properly. Evidently, it is of great importance to learn from the DAG structure for designing an ideal scheduling algorithm. On the contrary, the traditional algorithms just enqueue different available tasks by a high-level attribute directly without taking advantage of dependencies fully.

\noindent\textbf{Heterogeneous environment.}
Executors within a large scale cluster or cloud environment are usually heterogeneous, which probably have different computing speed, memory size, and transmission bandwidth. Furthermore, the difference in computing speed may cause a huge difference in computation time when executors deal with the same task. So allocating the task to a suitable executor is able to reduce execution time. Certainly, a naive idea is allocating all tasks to the fastest executor to get minimum time of task execution. However, the allocation strategy damages the parallelism of all task execution process. Although it obtains minimum execution time of a single task. In addition, the execution time of two similar tasks executed by two slow executors parallel is not less than executed by a fast executor when the processing speed of the fast executor is twice the slow executor. Therefore, an ideal scheduling algorithm should consider not only the parallelism of execution but also the difference in processing speed. In other words, the heterogeneity of executors makes a huge room for improvement of efficiency of DAG scheduling systems. Evidently, the task should be assigned to a specified executor which is more complicated than assignment in a homogeneous environment.


\noindent\textbf{Continuous arrival of jobs.} In addition to understanding task dependencies and executor heterogeneity, an ideal DAG scheduling algorithm should also deal with the situation that many jobs arrive in the system continuously. For the data processing platform, the jobs are submitted by different clients randomly. Hence, the DAG scheduling algorithm should schedule the tasks to executors continuously and plan ahead for the future arrival jobs. The future jobs arriving at the platform will change the current DAGs' structure. On the contrary, some existing algorithms only handle the situation where all jobs arrive at the system at start time. They can optimize the schedule result iteratively and modify the existing assignment gradually due to the changeless DAGs. Therefore, the DAG scheduling algorithm in continuous environment should make an assignment immediately which would be executed in time. Besides, every assignment decision should be completed in a short period of time to avoid blocking the execution of tasks.

\section{The proposed framework}\label{design}
This section presents the details of our proposed \lachesis. Specifically, \lachesisquad is composed of two parts: a GNN and a heuristic algorithm, which is corresponding to the two phases of the scheduling process, namely the node selection phase and the executor allocation phase. In the node selection phase, \lachesisquad uses a GNN to learn the complex relationship no matter tasks in the same job or in different jobs. In the executor allocation phase, \lachesisquad uses a heuristic search algorithm to assign the selected task to a suitable executor and decide to duplicate the corresponding task or not. Finally, we adopt a reinforcement learning paradigm to train \lachesisquad as an entirety. As scheduling DAG jobs, \lachesisquad repeats the above process until all tasks are assigned to executors. In particular, this kind of algorithm framework combine the advantages of RL and heuristic method to ensure high decision quality and fast inference, which can handle scalable task information and heterogeneous executor states.



\subsection{GNN Design}\label{sec:node_embedding}

The GNN contained in the \lachesisquad is mainly constructed by two parts, namely information embedding and a fully connected feed-forward network. In this section, we will focus on the detail of information embedding. Since we adopt a reinforcement learning paradigm to train \lachesis, the information of jobs in the system should be properly transferred to the \lachesisquad to function as state information. Obviously, the most naive idea is to construct a flat vector with all features of a job to represent the job in the system. However, a flat vector is incapable to represent the complicated relationship between different job stages. Hence, a information embedding process is leveraged to generate a representation consisting of all the useful information. 

\begin{figure}[t!]
 \centering
 \includegraphics[width=1.0\linewidth]{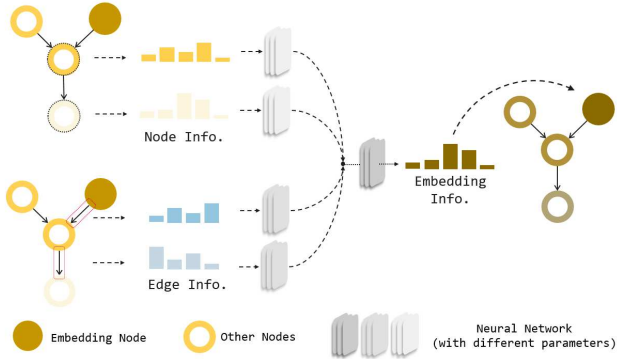}
 \caption{
Node embedding in a DAG job. For a certain embedding node, the information contained in its children nodes and edges connecting them is aggregated and transferred in a two-level neural network.
 }
 \label{fig:node embedding}
\end{figure}

Before concentrating on the process of information embedding, it is significant to choose useful features for a job stage as raw information. Evidently, good features of nodes may help the model represent DAG information clearly. And some high-level features such as $rank_{up}$ and $rank_{down}$ can represent the DAG information about the task, which are beneficial to the GNN to process the problem structure. For instance, $rank_{up}$ and $rank_{down}$, which can be calculated as Eq. (\ref{eq:rank_up}) and (\ref{eq:rank_down}) respectively, represent the length from the node to exit node (the node without children nodes) and the length from the task to the entry node (the node without parent nodes)~\cite{cpop_heft} respectively. In other words, the two features indicate the approximate position of the belonging DAG for the corresponding node. In particular, the job attributes contain the number of left tasks and the sum of average execution time for left tasks. Moreover, the node features contains all the features from the corresponding job such as the average left execution time, the average time cost of incoming data, the average time cost of outgoing data, $rank_{up}$, and $rank_{down}$.
\begin{equation}
  \label{eq:rank_up}
  \operatorname{rank}_{up}\left(n_{i}\right)=\frac{w_{i}}{\bar{v}}+\max _{n_{j} \in \xi\left(n_{i}\right)}\left(\frac{e_{i, j}}{c_{i j}}+\operatorname{rank}_{u p}\left(n_{j}\right)\right)
\end{equation}
\begin{equation}
  \label{eq:rank_down}
  \operatorname{rank}_{\text {down}}\left(n_{i}\right)=\max _{n_{j} \in \operatorname{\varphi}\left(n_{i}\right)}\left(\operatorname{rank}_{\text {down}}\left(n_{j}\right)+\frac{w_{j}}{v_{j}}+\frac{\overline{e_{j i}}}{c_{j i}}\right)
\end{equation}
In Eq. (\ref{eq:rank_up}), $\overline{v}$ means the average speed of all executors, and $e_{i j}$ indicates the data size transferred from node $n_i$ to node $n_j$. Generally, the $rank_{up}$ value of node $n_i$ equals the average execution time of itself adding the maximum value of the sum of communication cost and the maximum $rank_{up}$ of children nodes. Hence, $rank_{up}$ value is the maximum execution time for the paths from the node to exit node. Similarly, $rank_{down}$ value represents the maximum execution time for the paths from the entry node to the corresponding node. The above process is shown in Figure~\ref{fig:node embedding}.

\begin{figure}[t!]
 \centering
 \includegraphics[width=0.83\linewidth]{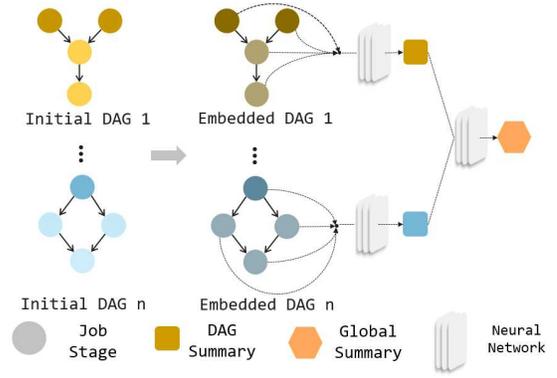}
 \caption{High-level summarization of DAG jobs' information. Nodes' information is pooled to generate DAG summaries. And all DAG summaries are passed through a neural network to generate a global summary.
 }
 \label{fig:information summary}
\end{figure}

The information embedding is composed of two procedures: node embedding and high-level summarization. In the node embedding procedure, the computation of every node embedding GNN layer can be expressed as follows:
\begin{equation}
  \label{eq:GNN}
  \mathbf{e}_{n}^{j}=g\left[\sum_{u \in \xi(n)} \left(\mathbf{e}_{u}^{j}\right)\right]+h\left[\sum_{u \in \gamma(n)} \left(\mathbf{a}_{u}^{j}\right)\right]+\mathbf{x}_{n}^{j}
\end{equation}
In Eq. (\ref{eq:GNN}), vector $x_{n}^j$ represents the node attributes corresponding to tasks in Job j. \lachesisquad constructs a per-node embedding $(G_j,x_n^j)\longmapsto e_n^j$. Eq. (\ref{eq:GNN}) illustrates the fundamental calculation operation of GNN of \lachesis. Similarly, the per-job embedding and global embedding can be calculated as Eq. (\ref{eq:GNN}). Node embedding leverage the connectivity of DAG to pool the information to each node. By stacking node embedding GNN layers together, a node can eventually incorporate information from across all the reachable nodes. However, node embedding is incapable to aggregate the information for a whole DAG job or a cluster of DAG jobs. Thus, we design the high-level summarization to tackle this flaw, as shown in Figure~\ref{fig:information summary}.




After this multi-layer data processing, we get the three different dimension representations of the jobs in the system. Next, those representations are computed through a fully connected feed forward network, outputting initial scores for candidate nodes. Then, a softmax layer is used to compute the probability of every node with the obtained scores. The probability of every node can be calculated as follows:
\begin{equation}
  \label{eq:softmax}
  P(n)=\frac{\exp \left(q_{n}^{i}\right)}{\sum_{u \in \mathcal{A}_{t}} \exp \left(q_{u}^{j(u)}\right)}
\end{equation}

\begin{figure}[t!]
 \centering
 \includegraphics[width=1.0\linewidth]{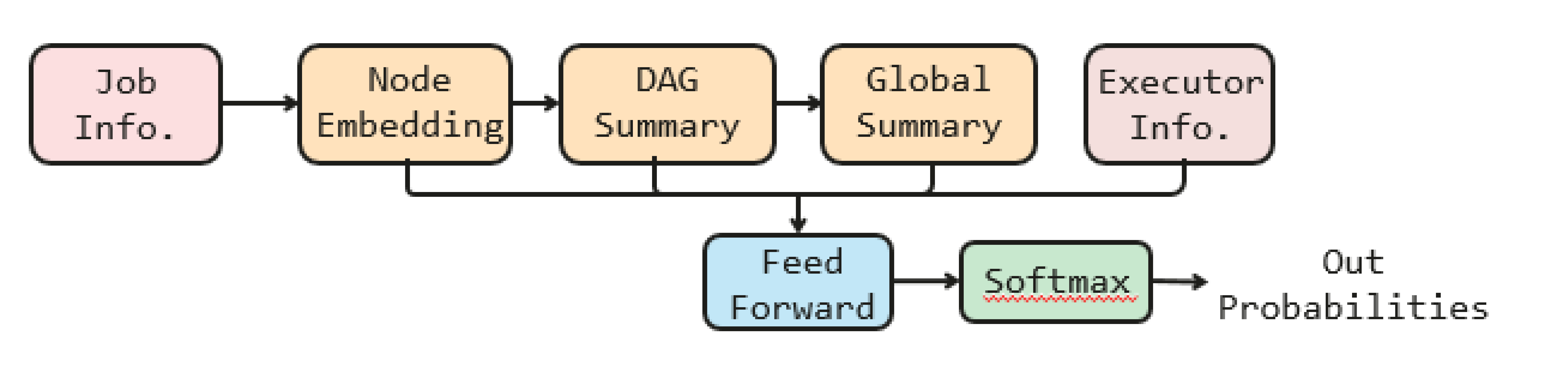}
 \caption{
Framework of the GNN used in \lachesis. Job information and executor information is the input of the GNN. Furthermore, the job information is incorporated to three different dimension representations. Finally, the GNN outputs the probabilities of each candidate job stage being selected.
 }
 \label{fig:whole_process_of_gnn}
\end{figure}

In Eq. (\ref{eq:softmax}), $P(n)$ represents the probability of the node $n$ and $q_n^i$ represents the score for the node $n$ in job $J_i$. And $\mathcal{A}_{t}$ indicates the executable node set for current state which is computed at every node selection step. All above procedures are depicted in Figure~\ref{fig:whole_process_of_gnn}.

After the process of GNN, the scheduling algorithm is able to select the next node based on current information which is a key part of the whole algorithm. Then, the selected node is supposed to be assigned to a suitable executor to minimize finish time of the node. Evidently, we can get a good schedule with a small makespan if we minimize the finish time of every node at every assignment step.

\subsection{Executor Allocation}\label{sec:resource_allocation}

Assigning a node to a suitable executor is another important step for the whole scheduling process. Obviously, a simple idea is making use of a neural network model to deal with node information and executor information to select a pair of nodes and executors at the same time. It handles the task and executor information together to generate an assignment pair which can be executed directly without other computation. However, the decision space of the neural network model would be the product of executable tasks and available executors which are too large to converge to a stable state. Therefore, we design a two-phase algorithm to cope with the problem which is capable of reducing the decision space, which includes the node selection phase and the executor allocation phase. In the first place, in the node selection phase, the algorithm only focuses on handling job information without considering executor information. In the second place, in the executor allocation phase, the model only makes decisions about executor information with the determinate selected node.

A heterogeneous environment makes the executor allocation more difficult than that in a homogeneous environment. Considering task duplication, task assignment should balance the computing cost and communication cost. In fact, heterogeneous executor information is standard and easy to deal with if the next node is determined. Therefore, we choose a heuristic search algorithm to allocate an executor for the selected node by considering execution efficiency. Task duplication should be considered under many conditions. Apparently, duplicating a task on different executors will probably reduce the earliest start time for the children tasks directly by saving communication cost. Besides, duplicating many tasks on different executors may also be beneficial to reduce the makespan in some cases. However, the online job processing procedure requires the algorithm to deal with the jobs as soon as possible. And the assignment generated by the algorithm is not changeable because it may be executed immediately. So it is not possible to optimize the assigned schedule through trying different duplication schedules.

%
Considering the efficiency of execution and the continuous execution model, we design the DEFT algorithm attempts to try to duplicate one parent node of the selected node to reduce the makespan avoiding duplicating multiple tasks simultaneously in at a single assignment step. In order to compute the earliest finish time of the selected node in task execution mode, copy parent earliest finish time algorithm (CPEFT) duplicates one parent of the select node on the executor before computing. In some cases, the earliest finish time in duplication execution mode is not less than that in common execution mode. So DEFT algorithm comprised of EFT and CPEFT is supposed to compare the result of EFT with that of CPEFT to obtain the minimum result. The CPEFT algorithm can be calculated as follows:

\begin{equation}
  \label{eq:aftc}
  AFTC(n_k,n_i,r_j)=\min _{r_{k} \in R_{n_{k}}}\left(A F T\left(n_{k}, r_{k}\right)+\frac{e_{k i}}{c_{k j}}\right)
\end{equation}


\begin{equation}
  \label{eq:cpeft}
  \resizebox{0.95\linewidth}{!}{
  $CPEFT \left(n_{p}, n_{i}, r_{j}\right)=
  \underset{n_{m} \in {\varphi}\left(n_{i}\right),n_{m} \neq n_{p}}{\max}
  \left[AFTC(n_p,n_i,r_j),AFTC(n_m,n_i,r_j)\right]+\frac{w_{i}}{v_{j}}$
  }
\end{equation}


In Eq. (\ref{eq:cpeft}), CPEFT attempts to duplicate each parent node of the selected node to calculate the earliest finish time and selects the minimum result. $R_{n_p}$ is the set of allocated executors for node $n_p$; $AFT$ and $AFTC$ are calculated as Eq. (\ref{eq:aft}) and Eq. (\ref{eq:aftc}). So DEFT can be calculated as follows:
\begin{equation}
  \label{eq:deft}
 \resizebox{0.90\linewidth}{!}{$
 DEFT\left(n_{i}\right)=\min \left[\min EFT\left(n_{i}, r_{k}\right),\underset{n_{p} \in \operatorname{\varphi}\left(n_{i}\right)}{\min }C P E F T\left(n_{p}, n_{i}, r_{j}\right)\right]$
}
\end{equation}

In Eq. (\ref{eq:deft}), the DEFT algorithm selects the minimum result in the results of EFT and CPEFT, which ensures that the result is the minimum one no mater duplicate the parent node or not. Appendix (Section~\ref{sec: supplementary materials}) provides the details of DEFT algorithm. Briefly, the heuristic algorithm DEFT searches all the executors and selects the earliest finish time assignment considering task duplication.

The heuristic algorithm makes the selected node matched with a suitable executor through two loops in a short time. In summary, we can calculate an assignment for a node through the node selection phase and the executor allocation phase.


\subsection{Algorithm Training}

Training the aforementioned algorithm model to suit the continuous execution environment is an intractable problem for \lachesis. \lachesisquad assigns a node to an executor at one step, and it repeats this procedure until all nodes are assigned. Once a node is assigned to an executor, the algorithm will decide the next node based on updated information about jobs and executors.
The timing to execute the scheduling algorithm in the continuous execution mode is significant. The scheduling algorithm should interact with the scheduling events because it can't schedule tasks all the time. In particular, the scheduling events happen when a job arrives at the system, a task is completed by executor and the system contains unassigned executable tasks. For each scheduling event, \lachesisquad probably schedules a task to the corresponding executor and updates information of the left tasks and executors.

RL training proceeds in episodes to optimize the objective. Each episode contains many scheduling events and actions generated by \lachesis. We use $T$ and $t_k$ to represent the number of actions in an episode which can vary in different episodes and the wall clock time for $k^{th}$ action respectively. Usually, the RL framework requires a reward to navigate the model to achieve the designed target. So after every action, \lachesisquad generates a reward of $r_k$ based on the high-level scheduling objective in the training process. For instance, \lachesisquad penalizes the agent $r_k = -(t_k-t_{k-1})$ after the $k^{th}$ action if the objective is minimizing the makespan. Minimizing the expected time-average of penalties $E[\frac{1}{t_{T}}\sum_{k=1}^{T}(t_k-t_{k-1})]$ is the target of the RL reward.

\lachesisquad uses synchronous actor-critic method~\cite{actor_critic_first} for RL training. Specifically, this method uses an actor network to choose the action and a critic network to score the potential selection actions, which is more stable than one network method for model training. In addition, all operations of \lachesisquad from GNN to the policy network are differentiable. We use the following elements to illustrate the RL model concisely.
 
\noindent\textbf{State (s)}: information about the system including the sum of average execution time of left tasks, the number of left tasks, average communication time, high-level features, etc.

\noindent\textbf{Action (a)}: the action of selecting the next task.

\noindent\textbf{Reward (r)}: $r_k=-(t_k-t_{k-1})$ which minimizes the makespan.

Provided that the agent collects the observation $(s_k, a_k, r_k)$ in the episode which has $T$ actions. \lachesisquad uses the actor-critic method to update the parameters $\theta$ of the policy $\pi_{\theta}(s_t,a_t)$ as follows:
 
 \begin{equation}
  \label{eq:update_parameter}
  \theta \leftarrow \theta+\alpha \nabla_{\theta} \log \pi_{\theta}\left(s_{k}, a_{k}\right) Q_{w}\left(s_{k}, a_{k}\right)
 \end{equation}

In Eq. (\ref{eq:update_parameter}), $\alpha$, $Q_{w}\left(s_{k}, a_{k}\right)$, and $ \log \pi_{\theta}\left(s_{k}, a_{k}\right)$ represent the learning rate, the score generated by the critic network for current $(s_k, a_k)$ and the various directions of the action $a_k$ on the state $s_k$ in the parameter space. Briefly, the goal of the equation is to get a better-than-average reward through changing the probability of the task being selected based on the score. The pseudo code in Algorithm~\ref{algo:training} provides more details about algorithm training.
\begin{algorithm}[t]
 \caption{Training process for \lachesisquad}
 \label{algo:training}
 \small
 \SetAlgoLined
 Initialize the weights of actor $\mu\left(s \mid \theta^{\mu}\right)$ $\pi_{\theta}$and critic $Q_{w}\left(s, a \mid \theta^{Q_{w}}\right)$

 \For{each iteration}
 {
 
  Sample a job arrival sequence randomly
 
  Sample episode length $\tau \sim exponential(\tau_{mean})$
 
  Receive initial observation $s_1$
 
  \For{k = 1 to $\tau$ }
  {
    Select action $a_k = \mu\left(s \mid \theta^{\mu}\right)$
   
    Execution action $a_k$ and observe reward $r_k$, state $s_{k+1}$
   
    Let $ y_k = r_k + \gamma Q_{w}^{\prime}(s_{k+1}, \mu^{\prime}(s_{k+1}|\theta^{\mu^{\prime}})|\theta^{Q_{w}^{\prime}})$
   
    Update critic by minimizing the loss: $L = {\frac{1}{N}} {\sum_{i}(y_i-Q_{w}(s_i,a_i|\theta^{Q_{w}}))^2}$
   
    Update the actor policy:
   
    $\left.\nabla_{\theta^{\mu} \mu}\right|_{_{\mathbf{a k}}}=\frac{1}{N} \sum_{\boldsymbol{k}} \nabla_{\theta^{\mu} \mu}\left(s \mid \theta^{\beta^{2}}\right) \nabla_{a} Q_{w}\left(s, a \mid \theta^{Q_{w}}\right)$
   
    Update the target networks:
     
     
      $ \Delta \theta \leftarrow \Delta \theta+\nabla_{\theta} \log \pi_{\theta}\left(s_{k}, a_{k}\right)Q_{w}(s_k,a_k)$
   
  }
 
  $ \tau_{mean} \leftarrow \tau_{mean} + \epsilon$
 
  $ \theta \leftarrow \theta + \alpha \Delta \theta$

 }
\end{algorithm}

%



\section{Evaluation}\label{evaluation}


\subsection{Implementation}\label{sec:implementation}


\begin{figure}[t!]
  \centering
  \includegraphics[width=0.80\linewidth]{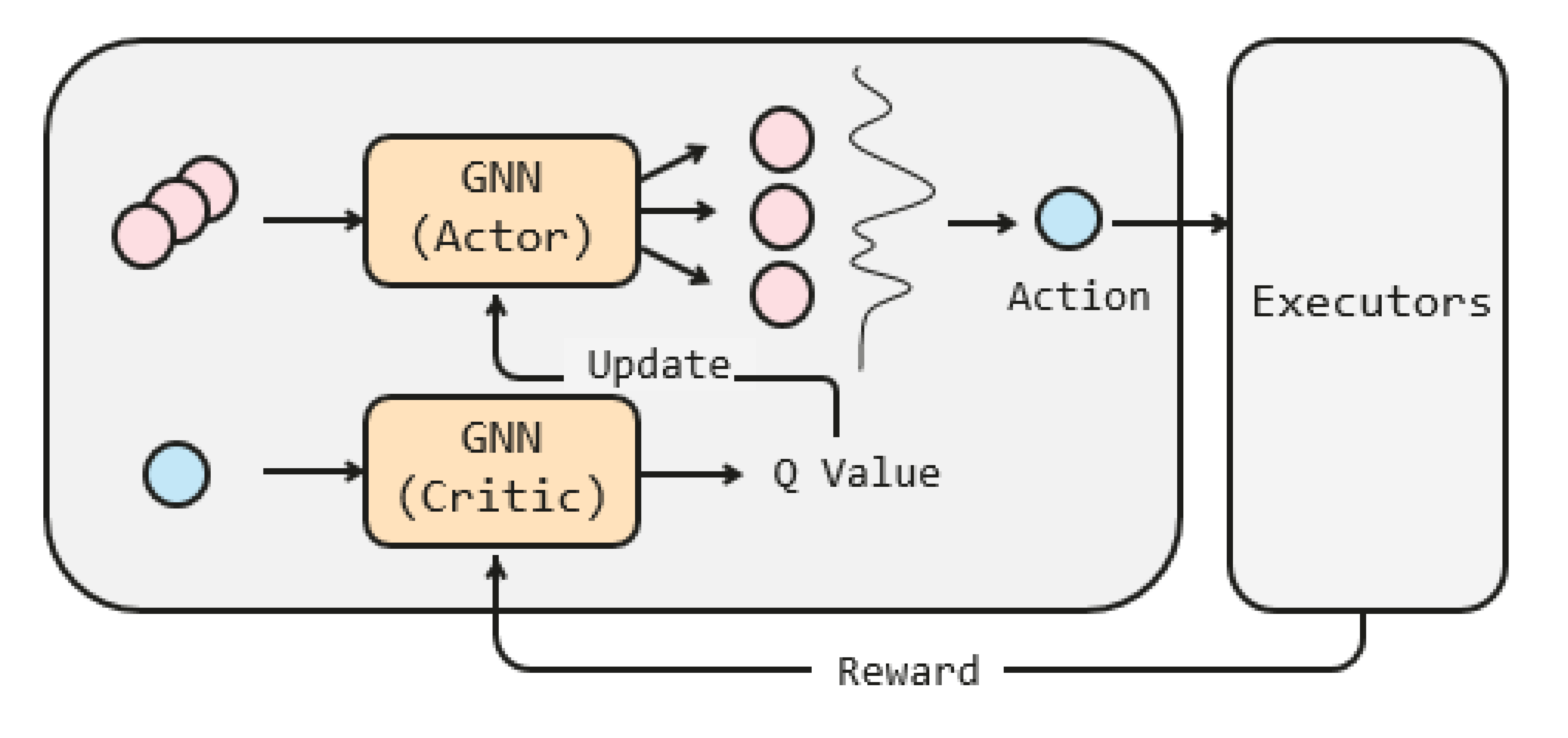}
  \caption{Actor-critic training framework of \lachesis.}
  \label{fig:actor-critic framework}
\end{figure}

We implement \lachesisquad as a plug-and-play scheduling service using Python, which can communicate with the data processing platform. Usually, the data processing platform consists of a master node and many resource nodes. The resource node executes the tasks dispatched by the master node and reports information about the resource node to the master node through the node manager which is an application in the resource node. Similarly, the resource manager dispatching tasks is an application of the master node.

The running procedure of the data processing platform with a \lachesisquad agent can be described as follows. First of all, clients submit jobs to the data processing platform. Then, the resource manager sends information about jobs and executors to \lachesisquad agent and gets an assignment for the next task from \lachesis. Next, the resource manager dispatches the task to the corresponding resource node based on the decisions of \lachesisquad agent. Particularly, the resource node reports the execution status to the master node with heartbeats. Therefore, \lachesisquad can be easily integrated into the data processing platform as a plug-and-play module. \lachesisquad agent only gets information from the master node and computes an assignment for the next task. The information that the master node transfers to \lachesisquad agent is state observation and resource status. State observation is described in Section~\ref{sec:node_embedding}. Resource status contains available execution time of the executor and the list of executed tasks on the resource node.




As for the detailed implementation of GNN in \lachesisquad, it consists of a three-layer modified graph convolution neural network which uses the technique of sharing parameters. Specifically, each layer only contains two non-linear function $f(\cdot)$ and $g(\cdot)$. Besides, the policy network contains three hidden fully connected layers with 32, 16, 8 units on each layer, respectively. As mentioned above, the RL neural network is a light-weight model which infers the next task less than 30 ms in most situations. For the heuristic method DEFT, it will cost $M$ computations if it does not duplicate the parent nodes where $M$ is the number of executors. Considering duplicating a parent node in every assignment step, it takes $P*M$ computations where $P$ is the number of parent tasks. So the complexity of the assignment one task to executor is $O(PM)$. For all tasks in the system, the complexity of the whole process is $O(EM)$ where $E$ is the number of edges.

A simulated environment for the data processing platform is beneficial to the offline training of \lachesis. Generally, training the RL model in a real environment occupies many executors with a low speed for training. Therefore, we designed a simulator for the data processing platform and trained our model on it. The details of the simulator are demonstrated in Appendix (Section~\ref{sec: supplementary materials}). The RL model trained with the simulator can be integrated with the true data processing platform well to generate good assignments for the system.

\subsection{Experiment Settings}



\noindent\textbf{Dataset.} The experiment dataset is generated from TPC-H workload~\cite{tpc_h}, which is a decision support benchmark. Specifically, the workload mainly contains TPC-H queries which are executed on a real data processing platform. In order to avoid processing the complex real environment, we extract the key information about task dependencies and workload size from the workload and evaluate \lachesisquad in the simulated environment with the TPC-H workload. Because the scheduling algorithm should be able to process jobs regardless of scale, we evaluate it with small scale jobs and large scale jobs. The jobs are selected from the TPC-H workload with six size types (2, 5, 10, 50, 80 and 100 GB) and 22 different job shapes. For the heterogeneous executors, we focus on processing speed of executors. So we collect the frequency of the Intel CPU range from 2.1 to 3.6 GHz. We simulate 50 executors in experiments and set computing speed from the frequency table randomly to realize the heterogeneity. In order to simplify the scheduling problem, we assume that transfer speed is the same between different executors.

\noindent\textbf{Compared baselines.} Seven strong baselines are compared with the proposed \lachesisquad:

1) First-In-First-Out (FIFO) scheduling method. It executes jobs based on an ascending order of arrival time that they arrive at the platform. And the algorithm uses the DEFT algorithm in the executor allocation phase.

2) Short Job First (SJF) scheduling method. It prioritizes the executable tasks by an ascending order of arrival time to generate the execution order. Similarly, the algorithm also allocates the selected task to an executor through DEFT algorithm.

3) HEFT algorithm. It is a high efficiency heuristic scheduling algorithm in non-duplication and batch mode. It prioritizes the tasks with high-level features $rank_{up}$ by a descending order and allocates executors with the EFT algorithm.

4) TDCA algorithm. It makes use of duplication and clustering techniques to minimize the makespan in batch mode.

5) A RL-based scheduling algorithm based Decima named Decima-DEFT. It combines the task selection method from Decima and the DEFT heuristic algorithm.

6) A high $rank_{up}$ first scheduling algorithm. It selects the next task based on the $rank_{up}$ feature. It also uses a DEFT algorithm to allocate executors for the selected task.

7) A high response ratio next (HRRN) algorithm. It selects the next task by the ratio $\frac{t_{wait}}{t_{wait}+t_{execution}}$ where $t_{wait}$ represents the wait time and $t_{execution}$ represents the execution time. DEFT is used for the executor allocation in the algorithm.

\begin{figure}[t!]
  \centering
  \includegraphics[width=0.7\linewidth]{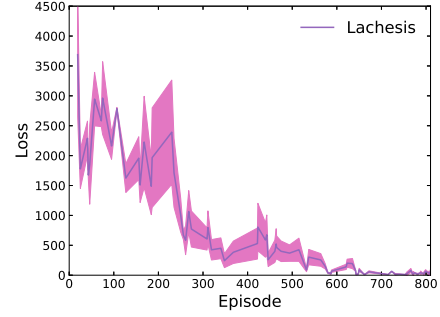}
  \caption{Learning curve of the proposed algorithm}
  \label{fig:act_loss_8500}
\end{figure}

\begin{figure*}[t]
  \centering
  \begin{subfigure}{0.24\textwidth}
    \centering
    \includegraphics[width=1\textwidth]{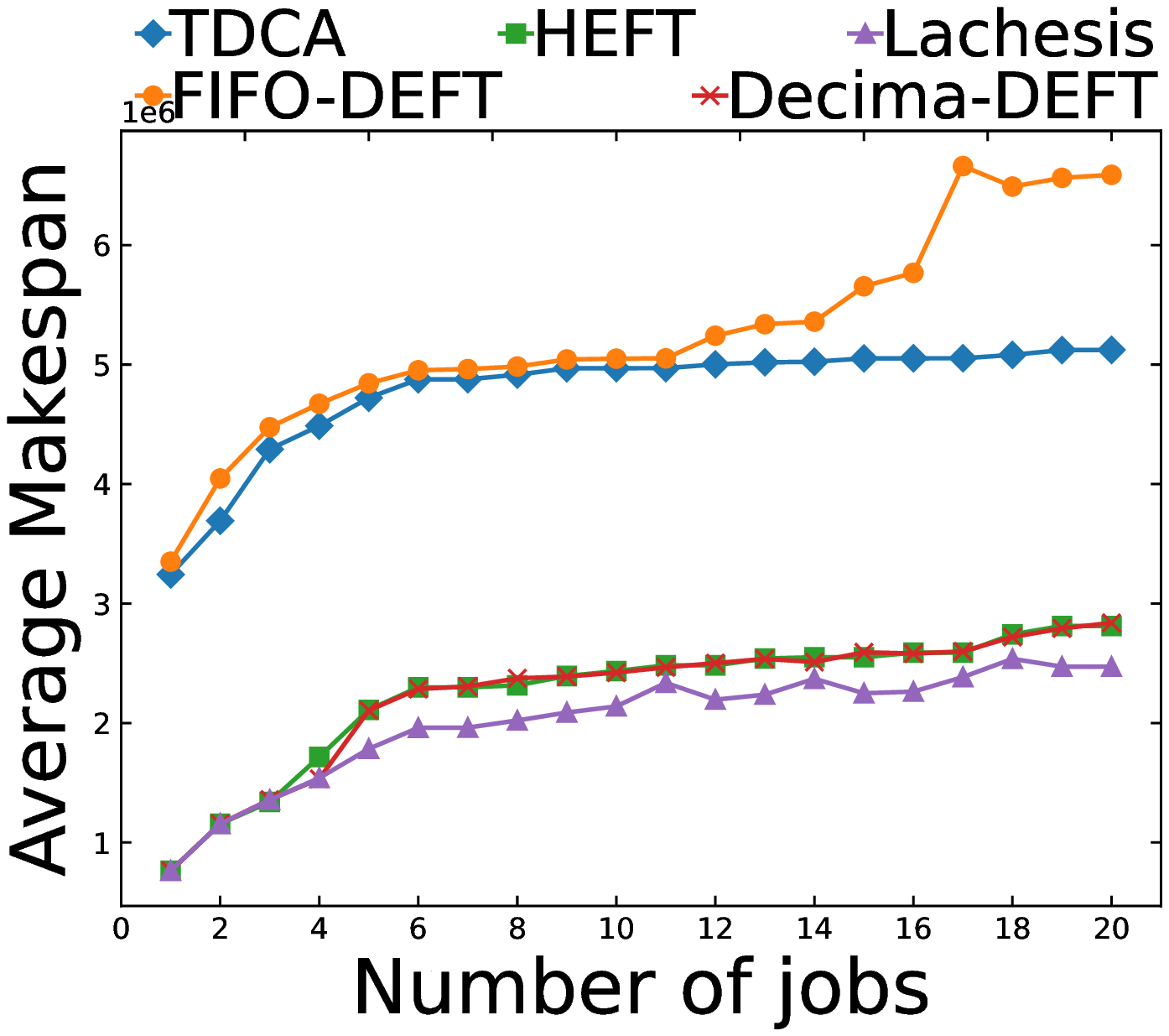}
    \caption{Average makespan}
    \label{fig:makespan_batch_small}
  \end{subfigure}
  \begin{subfigure}{0.24\textwidth}
    \centering
    \includegraphics[width=1\textwidth]{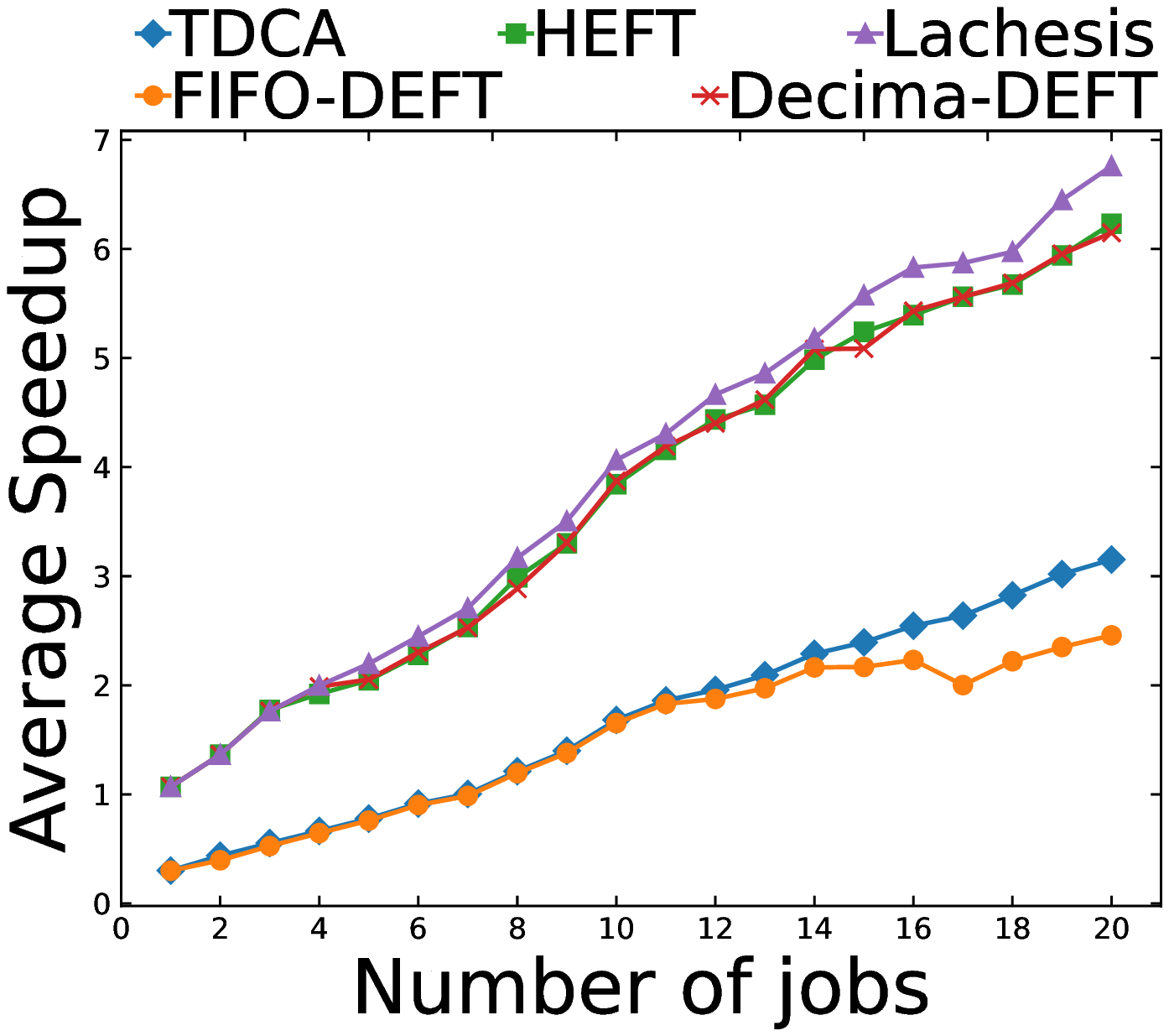}
    \caption{Average speedup}
    \label{fig:speedup_batch_small}
  \end{subfigure}
    \begin{subfigure}{0.24\textwidth}
    \centering
    \includegraphics[width=1\textwidth]{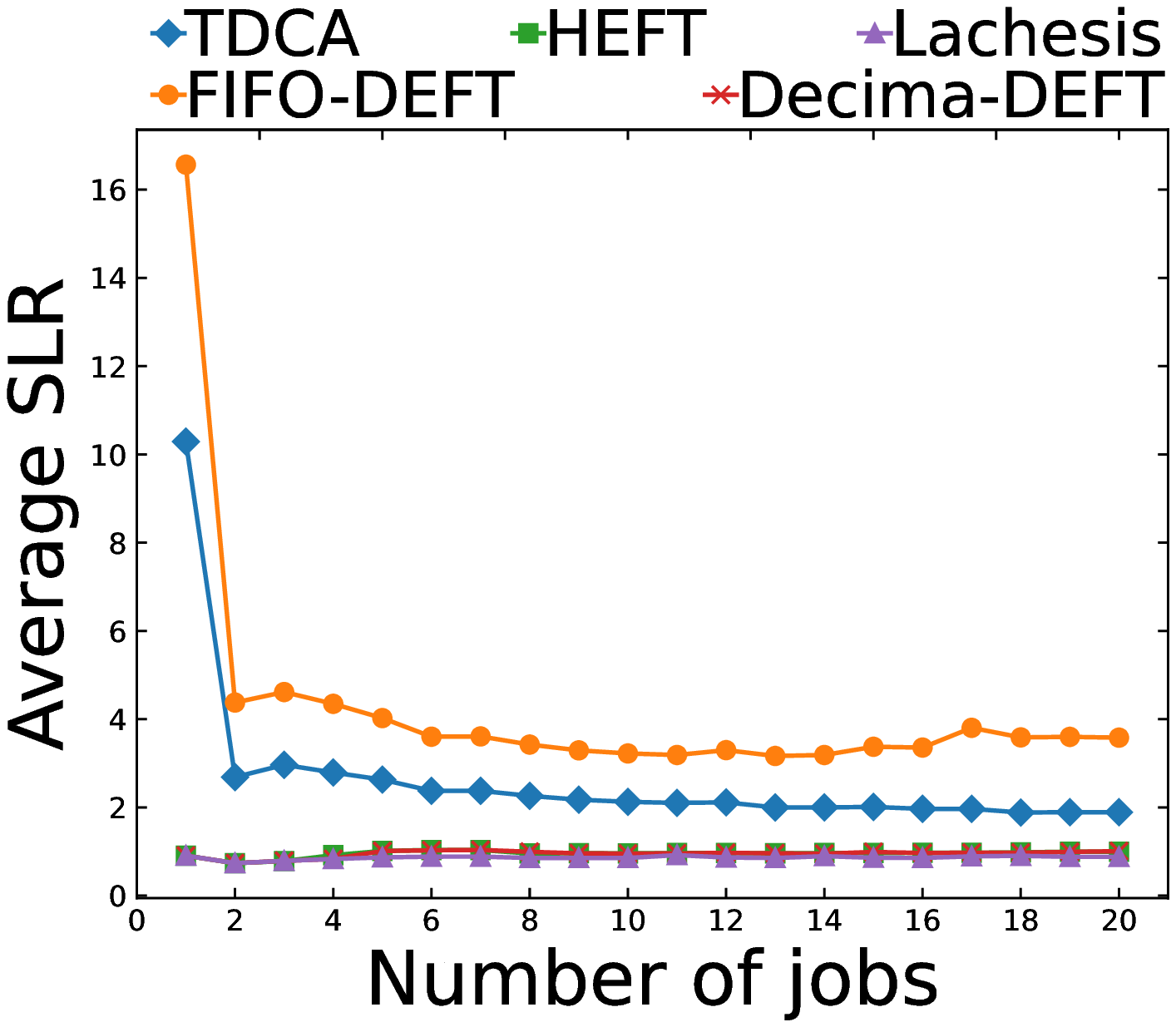}
    \caption{Average SLR}
    \label{fig:slr_batch_small}
  \end{subfigure}
\begin{subfigure}{0.24\textwidth}
    \centering
    \includegraphics[width=1\textwidth]{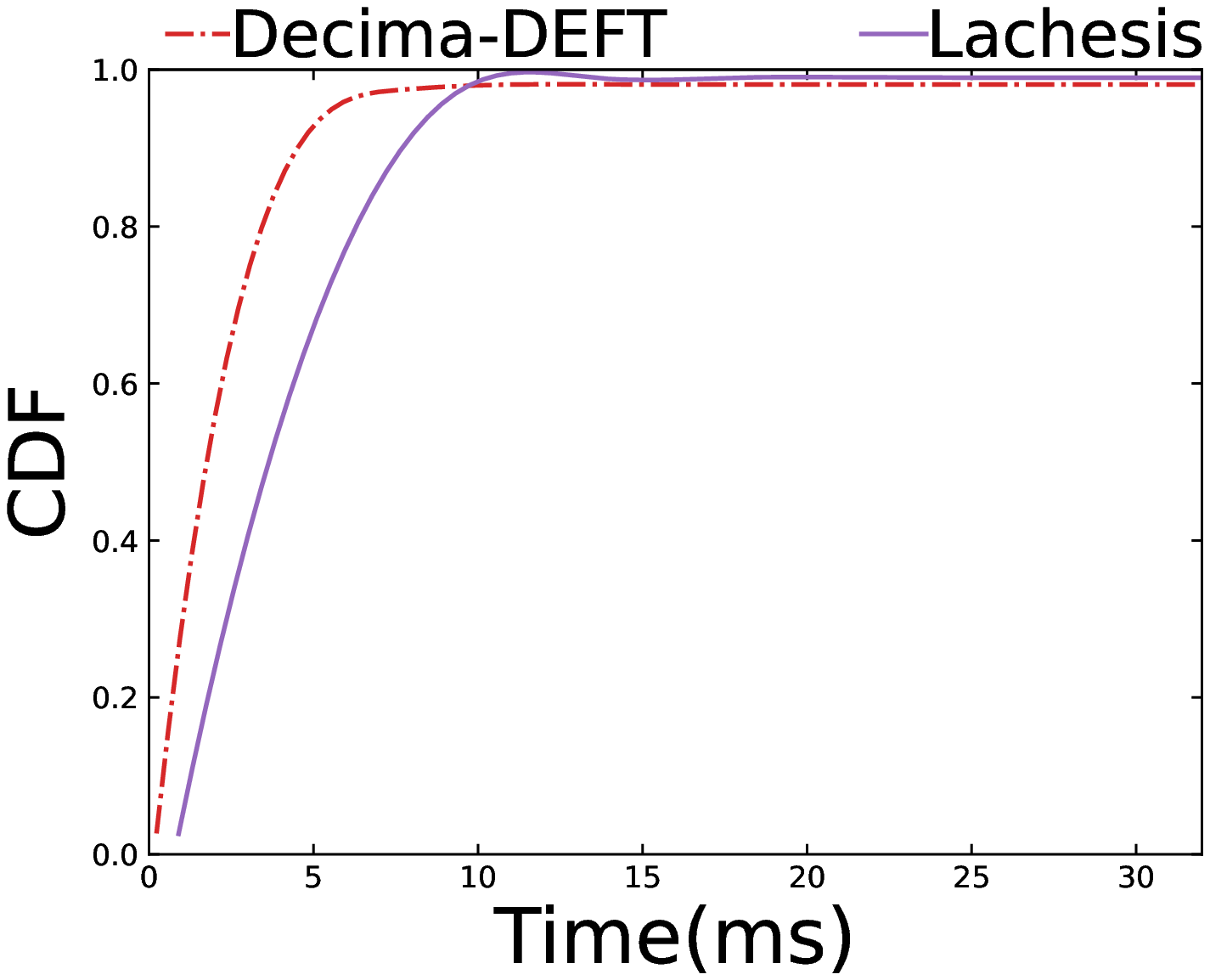}
    \caption{Decision time(ms)}
    \label{fig:time_batch_small}
  \end{subfigure}
  \caption{Experimental results under batch mode on small scale jobs}
  \label{fig:batch_small}
\end{figure*}

\begin{figure*}[t]
  \centering
  \begin{subfigure}{0.24\textwidth}
    \centering
    \includegraphics[width=1\textwidth]{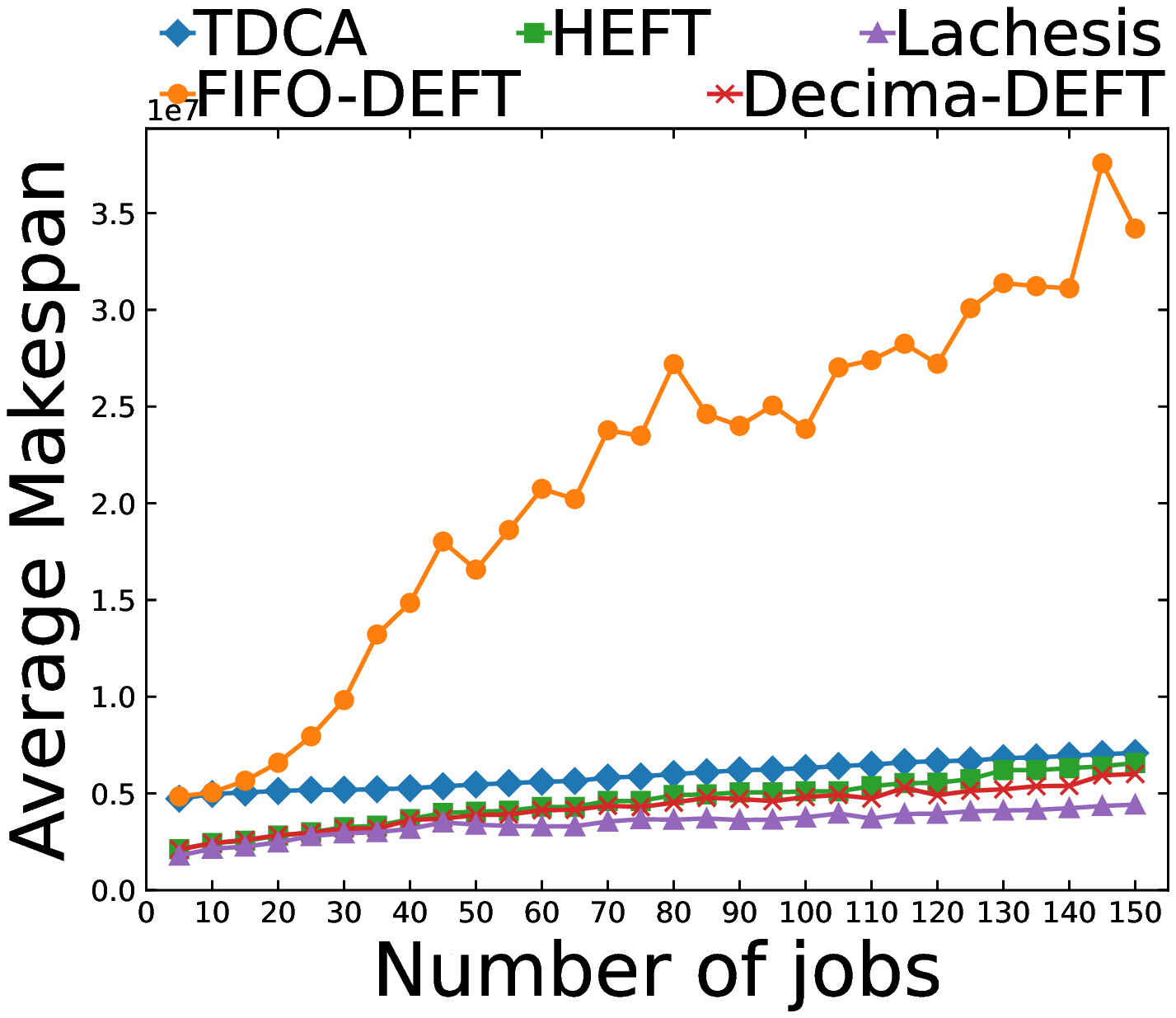}
    \caption{Average makespan}
    \label{fig:makespan_batch_large}
  \end{subfigure}
  \begin{subfigure}{0.24\textwidth}
    \centering
    \includegraphics[width=1\textwidth]{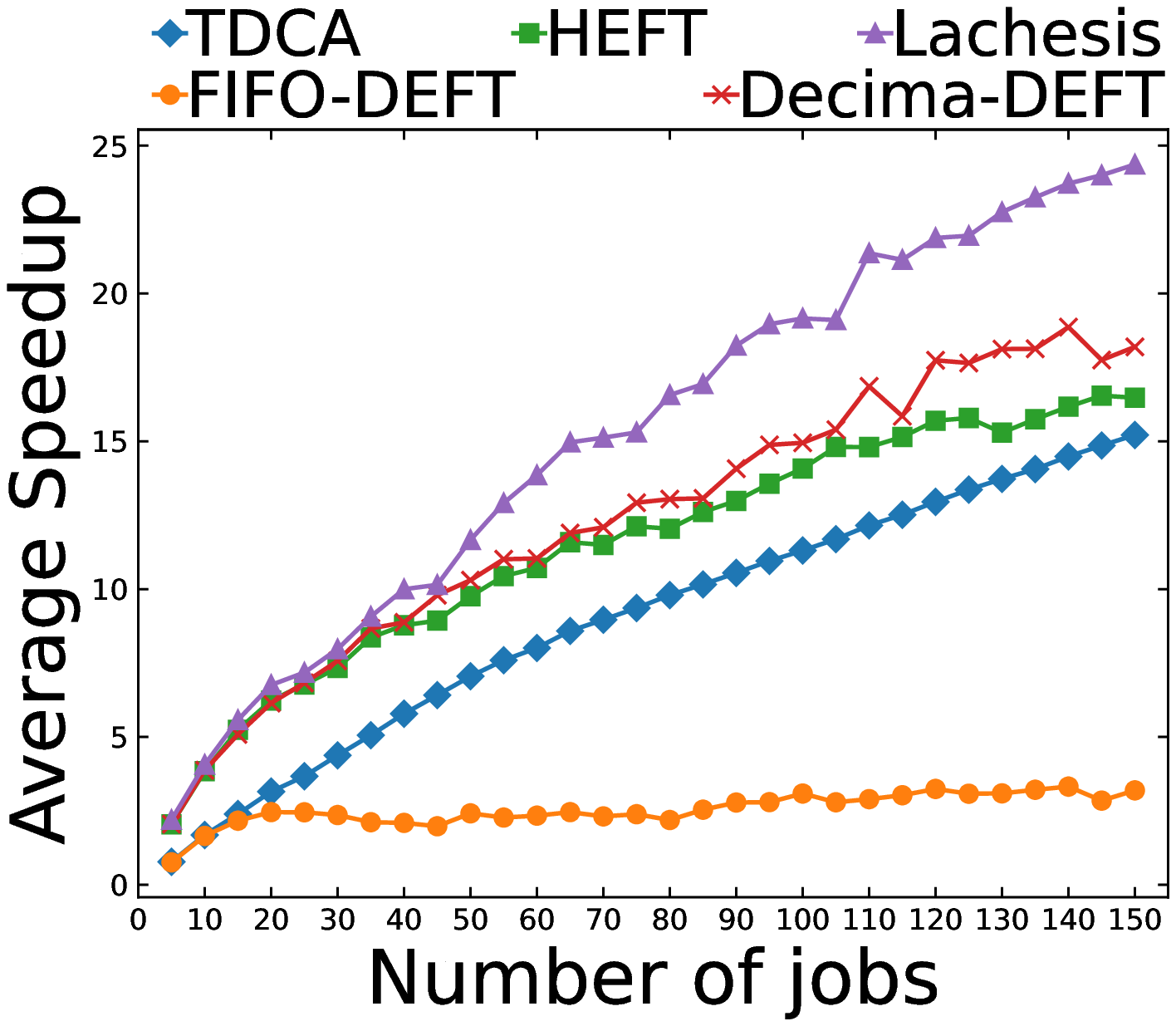}
    \caption{Average speedup}
    \label{fig:speedup_batch_large}
  \end{subfigure}
    \begin{subfigure}{0.24\textwidth}
    \centering
    \includegraphics[width=1\textwidth]{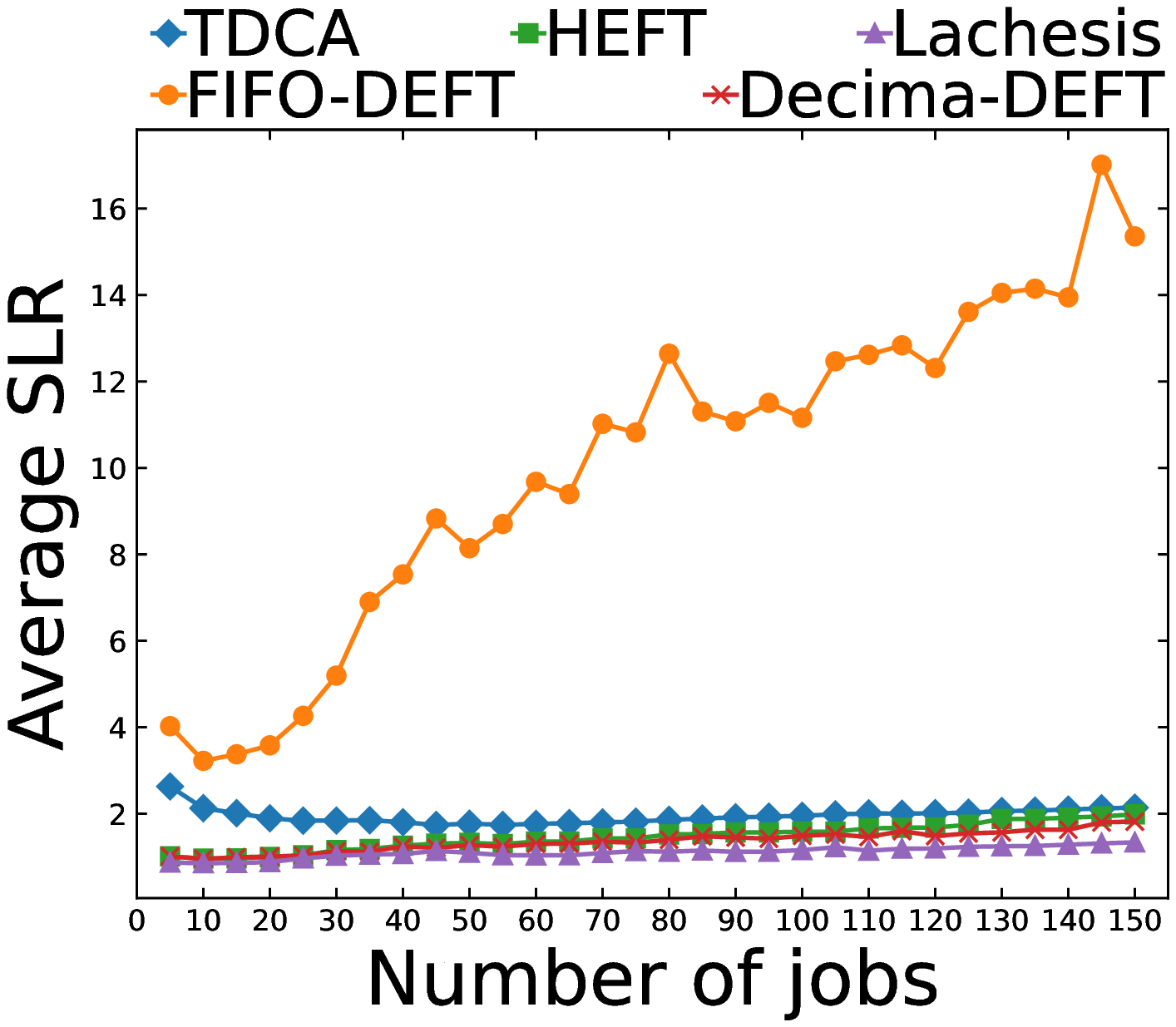}
    \caption{Average SLR}
    \label{fig:slr_batch_large}
  \end{subfigure}
  \begin{subfigure}{0.24\textwidth}
    \centering
    \includegraphics[width=1\textwidth]{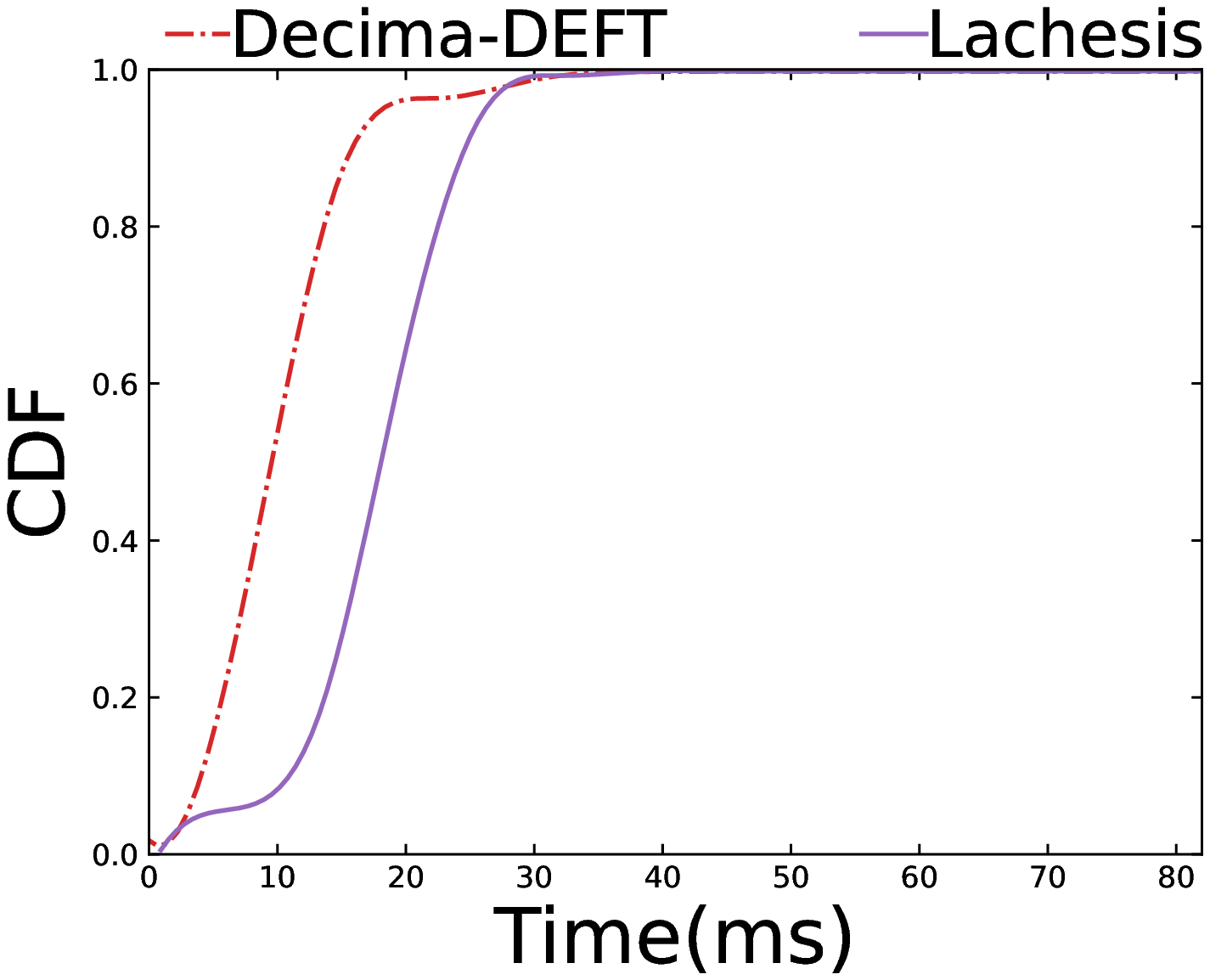}
    \caption{Decision time(ms)}
    \label{fig:time_batch_large}
  \end{subfigure}
  \caption{Experimental results under batch mode on large scale jobs}
  \label{fig:batch_large}
\end{figure*}

\noindent\textbf{Metrics.} We use makespan, speedup, and schedule length ratio to measure the performance of the scheduling algorithms. Evidently, different metrics can be used for measuring the performance in different angles. Specifically, makespan is the completion time of all jobs, and the speedup value is sequential execution time divided by actual execution time in batch mode. The sequential execution time is computed by assigning all tasks to the fastest executor. The actual execution time is the time of the generated schedule, also called makespan. The speedup value can be computed as follows:

\begin{equation}
    \label{eq:speed_up}
    Speedup=\frac{\min_{p_{j} \in Q}\left\{\sum_{n_{i} \in N} w_{i, j}\right\}}{ makespan }
\end{equation}
In Eq. (\ref{eq:speed_up}), the numerator is a determinate value for the same executor environment. Therefore, higher speedup value represents less actual execution time which is a better schedule result. Average speedup value is the average result of speedup value, which is used in our experiment to measure the experiment result.

It is necessary to normalize the schedule to compare the performance between different jobs. And the schedule length ratio (SLR) is the makespan of the actual schedule divided by the lower bound execution time, which can normalize the scheduling result. The SLR value of a schedule on the jobs can be defined as follows:
\begin{equation}
    \label{eq:slr}
    SLR=\frac{ makespan }{\sum_{n_{i} \in C P_{\mathrm{MIN}}} \min _{p_{j} \in Q}\left\{w_{i, j}\right\}}
\end{equation}

In Eq. (\ref{eq:slr}), the denominator means the sum of execution time for allocating nodes on the critical path ($CP_{min}$) to the fastest executor. And the critical path of a DAG is the minimum length of entry node to exit node. Besides, the denominator of the fraction only depends on the DAGs and executors without the influence of the scheduling algorithm, which is the lower bound of the problem. Evidently, the scheduling algorithm gets the lowest SLR result in a special situation indicating that it is the best algorithm with respect to performance. In our experiment, the average SLR is used to measure the scheduling algorithm over different jobs.

\subsection{Experimental Results}

\subsubsection{Convergence}

Figure~\ref{fig:act_loss_8500} shows the loss value of \lachesisquad decreases as episode increases on batch execution mode. \lachesisquad's loss converges to a small value gradually after 800 episodes. Every training episode only takes less than 100 ms to handle with jobs arrived at the platform in the beginning, which indicates the high efficiencies of the simulator.

\subsubsection{Batch mode}

Here we measure the schedules generated by \lachesisquad in batch mode using the aforementioned metrics. In the batch mode, jobs arrive at the system in the beginning and no jobs arrive in the future. The number of jobs in the system ranges from 1 to 20 in a small scale experiment. Besides, we select ten workloads for every different number of jobs to examine the algorithms in order to reduce the inference of randomness. 

Figure \ref{fig:makespan_batch_small} shows that the average makespan of \lachesisquad is less than the other four baseline scheduling algorithms. Moreover, \lachesisquad can reduce the average makespan of the best baseline algorithm by 15.3\% at most. The performance of HEFT is close to that of Decima-DEFT, because the duplication execution mode benefits a little for the makespan in small scale situations. Furthermore, the makespan of FIFO-DEFT is the maximum result in five comparing algorithms because the algorithm only selects the next node by the arrival sequence, which can't utilize the dependencies between different tasks. Noticeably, the result of TDCA is only better than that of FIFO-DEFT indicating that TDCA may be not suitable for this kind of job. 

Figure \ref{fig:speedup_batch_small} shows that speedup value of \lachesisquad is the maximum value within the five algorithms. This figure demonstrates that \lachesisquad is able to speed up the execution process up to 6.8 times compared to the sequential execution process. The speedup value ascends as the number of jobs ascends, which demonstrates that the more executors are used for execution tasks. Figure \ref{fig:slr_batch_small} reveals that the SLR Value of HEFT, Decima-DEFT and \lachesisquad are close to 1 and the value of \lachesisquad is the minimum value in five algorithms. The SLR value of FIFO-DEFT and TDCA decreases a lot when the number of jobs changes from 1 to 2 because the time of the critical path increases rapidly. Figure \ref{fig:time_batch_small} demonstrates that 98\% decision operation time of \lachesisquad is less than 14ms, although it is about 3ms bigger than that of Decima-DEFT algorithm.

Figure \ref{fig:makespan_batch_large} shows that \lachesisquad is also the best scheduling algorithm within five scheduling algorithms. Furthermore, \lachesisquad reduces the makespan of the second algorithm by 26.7\% at most, which shows the advantage of \lachesis. We can notice that Decima-DEFT is better than HEFT in most cases in our experiment. Maybe the reason for this situation is that the duplication method of Decima-DEFT plays an important role in reducing the makespan. In addition, FIFO-DEFT is also the worst scheduling algorithm in large scale experiment because it only adopts the arrival order. 

Figure \ref{fig:speedup_batch_large} shows that the speedup value of \lachesisquad exceeds the other four algorithms substantially. Especially, the speedup value of \lachesisquad exceeds the second algorithm by 35.2\% at most. Besides, the speedup value of FIFO-DEFT is almost unchanged after the number of jobs is bigger than 20 due to the unreasonable makespan of the algorithm. Similarly, Figure \ref{fig:slr_batch_large} indicates that \lachesisquad is the minimum value within all value of five algorithms. The SLR value of \lachesisquad and that of Decima-DEFT are between 1 and 2 indicating that the learning method is good for DAGs scheduling situations. The SLR value of TDCA ascends with the number of jobs ascends nearly because the critical path of the special job is almost unchanged and the makespan of TDCA increases. Figure \ref{fig:time_batch_large} illustrates that 98\% decision operations of \lachesisquad can be completed in 30 ms in batch mode of large scale. Obviously, the decision time of two learning algorithms increases with the number of jobs in the system. \lachesisquad is about 10ms slower than Decima-DEFT algorithm.


\subsubsection{Continuous mode}
\begin{figure}[!btp]
  \centering
  \begin{subfigure}{0.236\textwidth}
    \centering
    \includegraphics[width=1\textwidth]{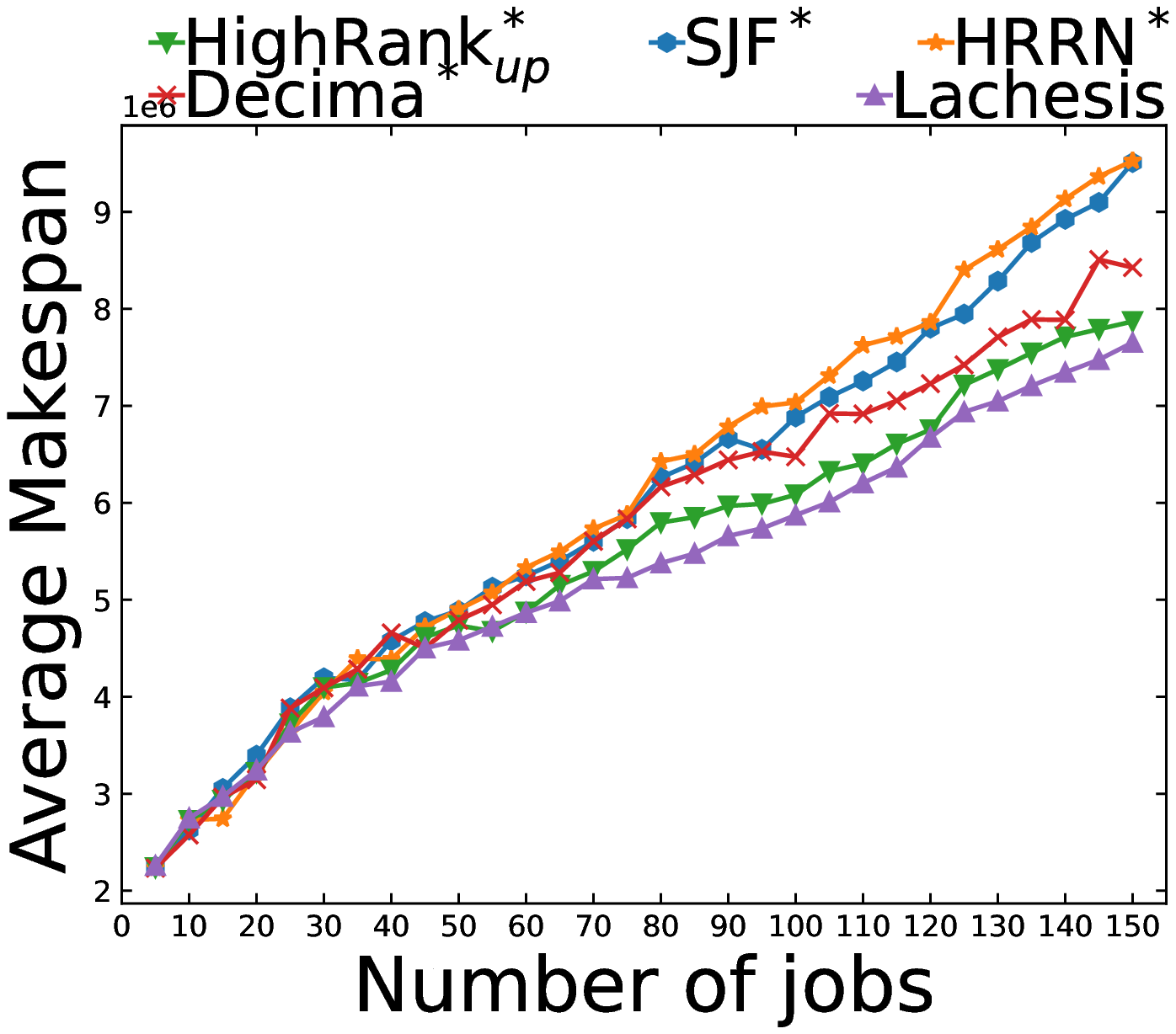}
    \caption{Average makespan}
    \label{fig:makespan_continuous}
  \end{subfigure}
  \begin{subfigure}{0.236\textwidth}
    \centering
    \includegraphics[width=1\textwidth]{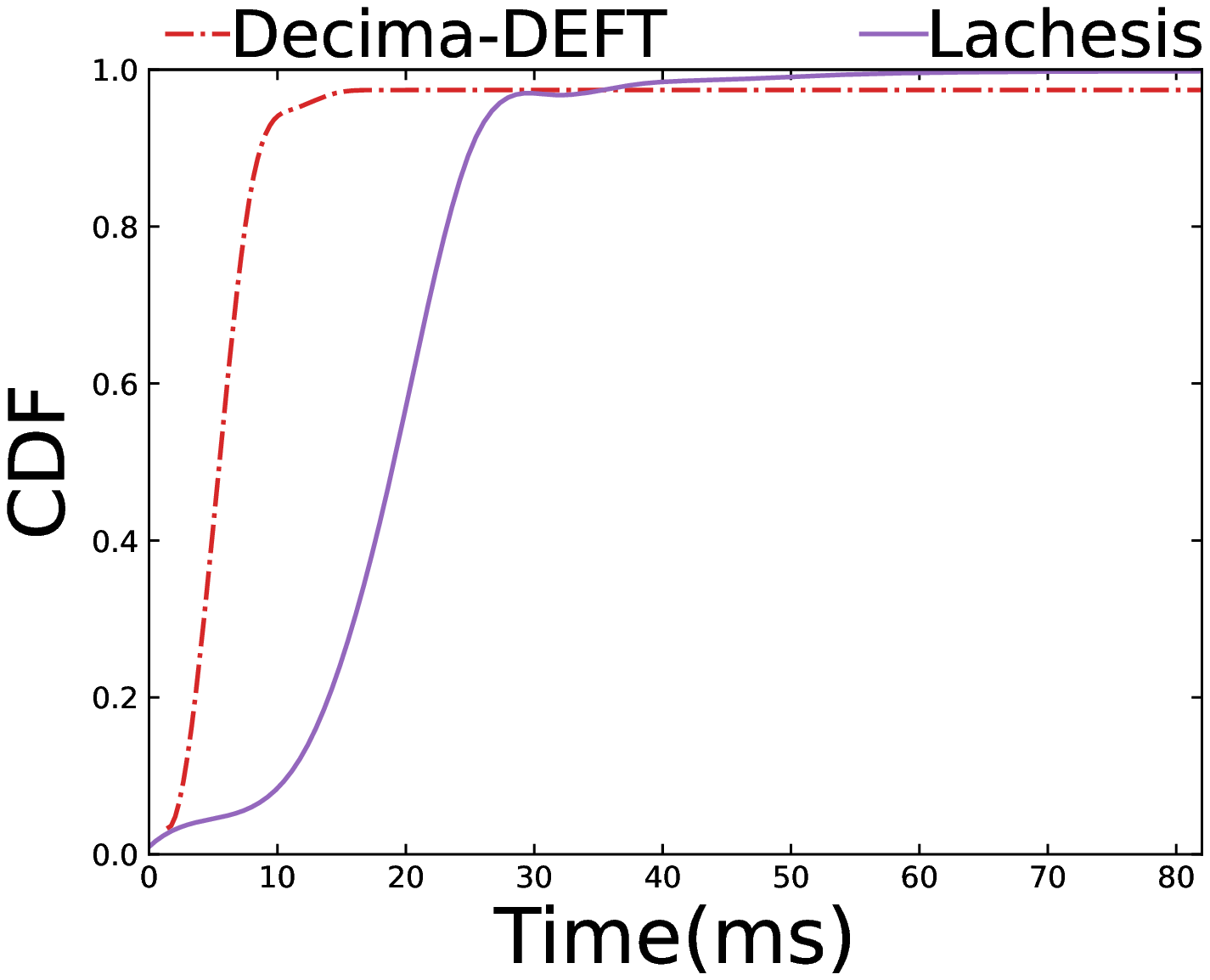}
    \caption{Decision time(ms)}
    \label{fig:time_continuous}
  \end{subfigure}
  \caption{Experimental results under continuous mode on large scale jobs}
  \label{fig:continuous_mode}
\end{figure}

In continuous mode, the jobs arrive at the system continuously and the scheduling algorithm needs to schedule online with current job information. Therefore, we let a job arrive at the system in the beginning and the others arrive at the system continuously. In order to simulate the arrival situation of data processing platform, we ensure that the arrival interval is the Poisson distribution with an average interval time of 45 seconds. 


Figure \ref{fig:makespan_continuous} shows that \lachesisquad is the best scheduling algorithm compared with other four baseline algorithms in most cases. In particular, SJF$^*$, HRRN$^*$, and HighRank$_{up}^*$ represents SJF-DEFT algorithm, HRRN-DEFT, and HighRank$_{up}$-DEFT respectively. Apparently, \lachesisquad is able to reduce the average makespan of the second algorithm by 7.4\% at most. Moreover, HighRank$_{up}$-DEFT is better than Decima-DEFT after the number of jobs is bigger than 50, which indicates that the node selection model of Decima-DEFT is worse than that of HighRank$_{up}$-DEFT. Besides, SJF-DEFT is better than the worst algorithm HRRN-DEFT representing that the shortest job first policy is better than the high respond ratio next policy in the node selection phase. For the data processing platform, scheduling decisions should be made in time to avoid unreasonable wait time. Figure \ref{fig:time_continuous} shows that \lachesisquad completes 98\% decision operation in 38ms. Hence, \lachesisquad is an efficient algorithm, although it is about 15ms slower than Decima-DEFT.

\section{Supplementary Materials}
\label{sec: supplementary materials}
Due to the limitation of space, we put part of contents in an external link, including the pseudo code of DEFT algorithm, implementation of the simulator for \lachesis, parameter/hyperparameter setting of referred algorithms, etc. Readers of interest can directly access the  \href{https://xijun-doc.oss-cn-hongkong.aliyuncs.com/MDM2022_submisson/Appendix.eps}{link}.

\section{Conclusion}\label{conclusion}
In this work, we propose a novel two-phase scheduling algorithm named \lachesisquad to address the job scheduling problem in heterogeneous environments. Specifically, it deal with node selection using reinforcement learning and copes with executor allocation by means of a sophisticated heuristic rules. The proposed \lachesisquad could be as a plug-and-play module in any kinds of data processing system which is supposed to cope with job scheduling problem. Extensive experimental results suggest that our proposed \lachesisquad greatly outperform the-state-of-the-art baselines in terms of the efficiency and effectiveness.




\bibliographystyle{IEEEtran}
\bibliography{reference}


\clearpage

\end{document}